\documentclass{article}

\usepackage{microtype}
\usepackage{graphicx}
\usepackage{subfigure}
\usepackage{booktabs} 
\usepackage[inline]{enumitem}
\usepackage{amsmath}
\usepackage{tikz}
\usepackage{xcolor}
\usepackage{graphicx}
\usepackage{svg}
\usepackage{algorithm}
\usepackage{algorithmic}
\usepackage{listings}
\usepackage{pythonhighlight}
\usepackage{etoolbox}

\usepackage{hyperref}

\usepackage[accepted]{mlsys2025}

\mlsystitlerunning{\papertitle}
\newcommand{\sys}{ETC}
\newcommand{\papertitle}{Event Tensor: A Unified Abstraction for Compiling Dynamic Megakernel}

\begin{document}

\twocolumn[
\mlsystitle{\papertitle}



\mlsyssetsymbol{equal}{*}
\mlsyssetsymbol{intern}{$\dagger$}

\begin{mlsysauthorlist}
\mlsysauthor{Hongyi Jin}{equal,cmu}
\mlsysauthor{Bohan Hou}{equal,cmu}
\mlsysauthor{Guanjie Wang}{equal,intern,sjtu}
\mlsysauthor{Ruihang Lai}{equal,cmu}
\mlsysauthor{Jinqi Chen}{nv}
\mlsysauthor{Zihao Ye}{nv}
\mlsysauthor{Yaxing Cai}{nv}
\mlsysauthor{Yixin Dong}{cmu}
\mlsysauthor{Xinhao Cheng}{cmu}
\mlsysauthor{Zhihao Zhang}{cmu}
\mlsysauthor{Yilong Zhao}{berkeley}
\mlsysauthor{Yingyi Huang}{nv}
\mlsysauthor{Lijie Yang}{princeton}
\mlsysauthor{Jinchen Jiang}{thu,intern}
\mlsysauthor{Gabriele Oliaro}{cmu}
\mlsysauthor{Jianan Ji}{cmu}
\mlsysauthor{Xupeng Miao}{pku}
\mlsysauthor{Vinod Grover}{nv}
\mlsysauthor{Todd C. Mowry}{cmu}
\mlsysauthor{Zhihao Jia}{cmu}
\mlsysauthor{Tianqi Chen}{cmu,nv}
\end{mlsysauthorlist}

\mlsysaffiliation{cmu}{Carnegie Mellon University}
\mlsysaffiliation{sjtu}{Shanghai Jiao Tong University}
\mlsysaffiliation{nv}{NVIDIA}
\mlsysaffiliation{berkeley}{University of California, Berkeley}
\mlsysaffiliation{princeton}{Princeton University}
\mlsysaffiliation{thu}{Tsinghua University}
\mlsysaffiliation{pku}{Peking University}

\mlsyscorrespondingauthor{Hongyi Jin}{hongyij@andrew.cmu.edu}

\mlsyskeywords{Machine Learning, MLSys}

\vskip 0.3in

\begin{abstract}
Modern GPU workloads, especially large language model (LLM) inference, suffer from kernel launch overheads and coarse synchronization that limit inter-kernel parallelism.
Recent megakernel techniques fuse multiple operators into a single persistent kernel to eliminate launch gaps and expose inter-kernel parallelism, but struggle to handle dynamic shapes and data-dependent computation in real workloads.
We present {\em Event Tensor}, a unified compiler abstraction for dynamic megakernels.
Event Tensor encodes dependencies between tiled tasks, and enables first-class support for both shape and data-dependent dynamism.
Built atop this abstraction, our Event Tensor Compiler (\sys{}) applies static and dynamic scheduling transformations to generate high-performance persistent kernels.
Evaluations show that \sys{} achieves state-of-the-art LLM serving latency while significantly reducing system warmup overhead.

\end{abstract}
]

\newcommand{\todo}[1]{\textcolor{red}{\textbf{TODO:} #1}}
\newcommand{\revised}[1]{#1}
\newcommand{\hongyi}[1]{\textcolor{orange}{hongyi: #1}}
\newcommand{\ruihang}[1]{\textcolor{violet}{Ruihang: #1}}
\newcommand{\yixin}[1]{\textcolor{violet}{Yixin: #1}}
\newcommand{\tqc}[1]{\textcolor{blue}{TQ: #1}}
\newcommand{\guanjie}[1]{\textcolor{violet}{Guanjie: #1}}
\newcommand{\zz}[1]{\textcolor{violet}{[ZZ: #1]}}
\newcommand{\hiddenNote}[1]{}

\newcommand{\calldevice}{\texttt{call\_device}}

\newcommand{\keypointcomment}[1]{}

\newcommand{\vspacebeforecap}{\vspace{-1em}}
\newcommand{\vspaceaftercap}{\vspace{-1em}}

\newcommand*\circled[1]{\tikz[baseline=(char.base)]{
            \node[shape=circle,draw,inner sep=0.8pt] (char) {#1};}}

\newcommand{\MyPara}[1]{\vspace{.0em}\noindent\textbf{#1}}

\newcommand{\squishlist}{
   \begin{list}{$\bullet$}
    { \setlength{\itemsep}{1pt}      \setlength{\parsep}{3pt}
      \setlength{\topsep}{3pt}       \setlength{\partopsep}{0pt}
      \setlength{\leftmargin}{1em} \setlength{\labelwidth}{1em}
      \setlength{\labelsep}{0.5em} } }

\newcommand{\squishlisttwo}{
   \begin{list}{$\bullet$}
    { \setlength{\itemsep}{0pt}    \setlength{\parsep}{0pt}
      \setlength{\topsep}{0pt}     \setlength{\partopsep}{0pt}
      \setlength{\leftmargin}{2em} \setlength{\labelwidth}{1.5em}
      \setlength{\labelsep}{0.5em} } }

\newcommand{\squishend}{
    \end{list}  }



\printAffiliationsAndNotice{\mlsysEqualContribution \textsuperscript{$\dagger$}Work done while at CMU. } 

\section{Introduction}


Efficient deployment of machine learning (ML) applications requires minimizing latency and maximizing hardware utilization, making system performance optimization~\cite{10.1145/3600006.3613165, zhu2025nanoflow, ye2025flashinfer, zhong2024distserve} a critical research frontier.
As GPUs continue to scale in speed and parallelism, several forms of system overhead in conventional GPU scheduling models have emerged as dominant bottlenecks that constrain end-to-end efficiency.

The first source of overhead arises from kernel launches. Current systems such as PyTorch~\cite{paszke2019pytorch} launch GPU kernels sequentially from the host CPU~(Figure~\ref{fig:scheduling-models}, upper left). During LLM inference, each auto-regressive decoding step may involve hundreds or even thousands of fine-grained operations, where the launch overhead cannot be effectively amortized. Each kernel launch typically incurs 5--10 $\mu$s of latency, while the fastest kernels may complete in 2 $\mu$s, making the launch overhead dominant.

The second source of overhead stems from kernel boundaries, which enforce implicit synchronization between consecutive kernels. In many cases, later kernels depend only on a subset of results from prior ones; in principle, these kernels could be overlapped or pipelined to improve throughput. However, the boundaries between kernels hinder such fine-grained inter-kernel parallelism, leaving significant performance on the table.

\begin{figure}[!t]
    \centering
    \includegraphics[width=0.45\textwidth]{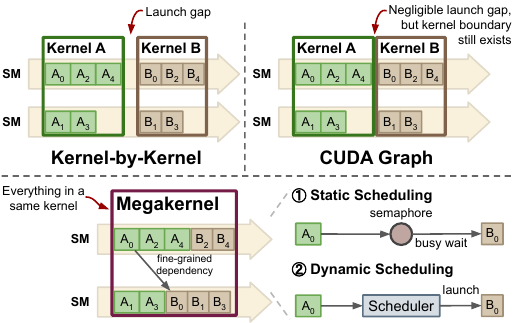} 
    \caption{Different GPU scheduling models. Kernel-by-kernel and CUDA Graph scheduling models enforce a coarse-grained sequential execution. Megakernels break operations into smaller tasks, achieving inter-kernel parallelism.
    }
    \label{fig:scheduling-models} 
    \vspaceaftercap
\end{figure}

\begin{figure*}[t!]
    \centering
    \includegraphics[width=0.93\textwidth]{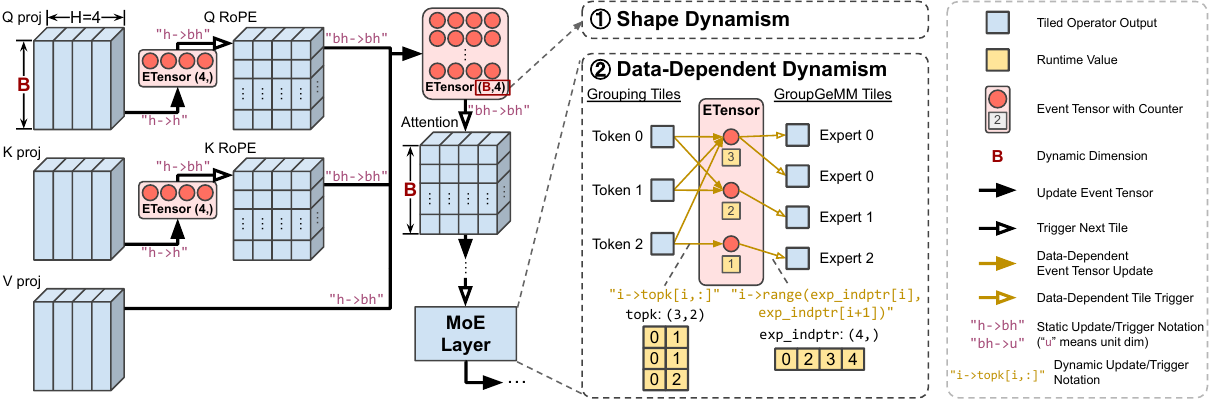}
    \vspacebeforecap
    \caption{Event Tensor abstraction overview. A computation graph (left) is partitioned into tiled operators (\emph{tasks}), and the Event Tensor captures fine-grained dependencies between tasks as a first-class, symbolic-shaped object, handling the primary sources of dynamism inherent to LLM serving: \circled{1} Shape Dynamism: Tiled tensors and Event Tensors have symbolic dimensions, such as the dynamic batch size \textbf{B}. \circled{2} Data-Dependent Dynamism: The MoE layer (right) details how dependencies are resolved at runtime. Data-dependent updates and triggers (yellow arrows) use runtime-computed values such as \texttt{topk} and \texttt{exp\_indptr} to dynamically manage the task execution. }
    \label{fig:overview} 
    \vspaceaftercap
\end{figure*}

Several recent efforts have attempted to partially address these limitations.
First, many systems adopt runtime techniques such as CUDA Graphs~\cite{Cudagraph}~(Figure~\ref{fig:scheduling-models}, upper right), which reduce kernel launch overhead by capturing and replaying a fixed sequence of kernels.
However, CUDA Graphs preserve kernel boundaries and thus cannot expose inter-kernel parallelism.


More recently, {\em megakernel} optimizations~\cite{tkmega,mpk} have emerged as a promising alternative~(Figure~\ref{fig:scheduling-models}, lower).
The key idea is to fuse multiple operators into a single persistent kernel, eliminating kernel launch overheads and enabling inter-kernel parallelism. Each operator is decomposed into fine-grained tiles of computation, or \emph{tasks}, which are distributed across streaming multiprocessors (SMs). The tasks and their dependencies form a task graph, whose execution is orchestrated through lightweight runtime signaling to preserve dependency while maximizing concurrency.

Despite its promise, deploying megakernels for LLM inference workloads remains challenging for two key reasons:


\MyPara{Dynamism challenges.}
Modern LLM serving workloads are inherently dynamic. With the introduction of continuous batching, the system must handle variable input shapes. Supporting such dynamic shapes within a single megakernel is challenging, as it often requires regenerating or recompiling the kernel for every possible shape. This can incur prohibitive startup latency or become impossible when the space of potential shapes is too large. Furthermore, models such as Mixture-of-Experts (MoE) introduce data-dependent control flow (e.g., expert routing), which requires dynamic tracking of fine-grained task dependencies to exploit inter-operator parallelism. Current approaches lack abstractions to express such fine-grained data dependency.
Notably, this challenge is not unique to megakernels---runtime dynamism also poses significant difficulties for conventional methods such as CUDA Graphs, where recapturing and managing CUDA Graphs across dynamic shapes is a major pain point for production LLM serving systems.
\revised{These dynamism challenges are particularly important for emerging latency-sensitive applications such as real-time agentic workflows and interactive coding assistants, where low-batch inference dominates and inter-kernel parallelism is critical for reducing per-request latency.}



\MyPara{Programmability challenges.}
Megakernel programming introduces substantial complexity. Developers must reason about complex, fine-grained dependencies among tasks, which are error-prone and difficult to maintain.
The shape and data-dependent dynamism further complicate this process.
Moreover, multiple task-management strategies may be desirable depending on the workload.
For instance, \emph{static scheduling} assigns each SM a predetermined queue of tasks before kernel launch, while \emph{dynamic scheduling} employs an on-GPU scheduler to dispatch ready tasks at runtime.
Ideally, developers should be able to seamlessly select or combine these scheduling strategies without reimplementing the entire kernel, enabling workload-specific scheduling.

In this paper we present \textbf{Event Tensor}~(\autoref{fig:overview}), an abstraction designed to simplify the compilation and execution of dynamic megakernels.
We define an \emph{event} as a primitive representing the completion of a set of tasks at the granularity of GPU SMs.
Because megakernels partition operators into a large number of tile-level tasks, the corresponding synchronization events naturally form multi-dimensional structures similar to data tensors.
An \emph{Event Tensor} is a multi-dimensional array of such events, providing a compact, first-class representation for fine-grained synchronization within a megakernel.
\revised{While semaphore-based synchronization is a well-known primitive, our core novelty lies in elevating these primitives into first-class tensors within the compiler IR.}
By unifying events into tensor form, Event Tensors leverage existing compiler support for symbolic shapes~\cite{PyTorch2,lai2025relax}, allowing tensor dimensions to remain symbolic and thereby compactly representing dynamic-shape computations.
Furthermore, Event Tensors express data-dependent dynamism through index expressions that map task coordinates to event coordinates.
\revised{Built upon this abstraction, \sys{} automatically transforms these dependencies into highly optimized, persistent megakernels, generalizing optimizations that previously required manual, specialized engineering.}





Building on the Event Tensor abstraction, we develop a systematic compiler pipeline that automatically fuses and schedules operators for inter-kernel parallelism. Starting from a computational graph annotated with explicit operators and Event Tensors,  the compiler applies a series of scheduling transformations to lower the program into an executable megakernel. These transformations support multiple scheduling strategies---ranging from fully static to dynamically load-balanced execution---each representing a different trade-off between synchronization overhead and runtime adaptability.
By unifying previously manual fusion and hand-crafted scheduling techniques~\cite{tkmega, mpk} into compiler transformations, \sys{} significantly reduces engineering effort to construct megakernels, while improving their runtime performance.

We evaluate \sys{} on a diverse set of LLM serving workloads, \revised{comparing against highly competitive, industry-level baselines (e.g., vLLM and SGLang) that already employ aggressive optimizations such as CUDA Graphs, Programmatic Dependent Launch (PDL), and {\tt torch.compile}.}
Our results show that \sys{} achieves substantial speedups over these systems while efficiently supporting both shape- and data-dependent dynamism, without requiring runtime graph recapture or recompilation.
For tensor-parallel workloads, our compiler-driven overlap of computation and communication achieves up to 1.40x speedup on fused GEMM and Reduce-Scatter kernels.
For data-dependent workloads such as MoE, our megakernels outperform specialized libraries by up to 1.23x.
In dynamic-shape, low-batch inference scenarios, \sys{} matches or exceeds the performance of these highly optimized inference systems, while reducing engine warm-up overheads by up to 3.5x.
\revised{Achieving even moderate speedups over such strong baselines translates to substantial economic value at datacenter scale.
Beyond raw speed, \sys{} achieves true ahead-of-time (AOT) compilation for dynamic workloads, completely eliminating runtime compilation overhead and the management complexity of repeated CUDA Graph recapture---a major pain point in production serving systems.
Furthermore, \sys{} automates megakernel fusion of complex, data-dependent subgraphs (e.g., MoE layers, GEMM+communication), significantly reducing programming complexity while remaining composable with existing serving engines.}
\sys{} has been incorporated into a major open-source system.
The simplicity and generality of the Event Tensor abstraction, together with the accompanying compiler framework, can benefit the broader machine learning systems and compiler community.


\section{Event Tensor Abstraction}


\subsection{Language Constructs}

\begin{figure}[t]
    \centering
    \includegraphics[width=0.45\textwidth]{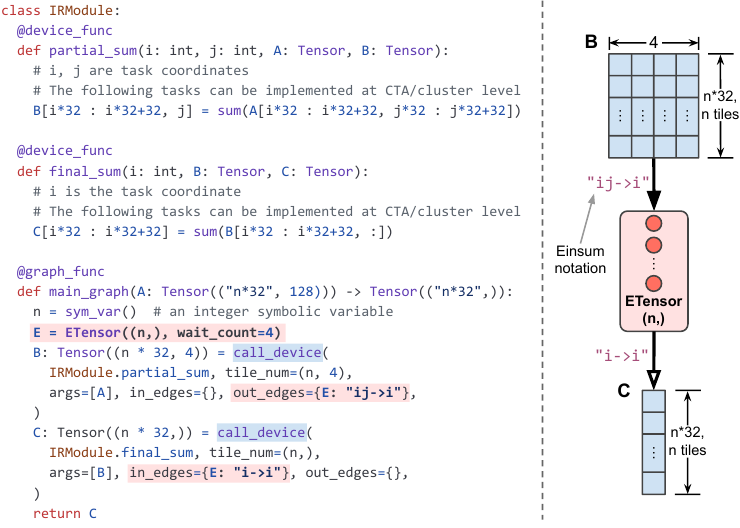} 
    \vspace{-1em}
    \caption{Example Event Tensor--based program.}
    \vspace{-2em}
    \label{fig:example_code} 
\end{figure}


\hiddenNote{Event Tensor with a conceptual example}

\hiddenNote{What is event tensor}



We first introduce the main language constructs in Event Tensor--based programs.

\textbf{Device Function.} A device function defines a grid of tasks launched in parallel on the GPU.
Each launch is parameterized by a multidimensional coordinate, where each coordinate identifies a task tile executed on a streaming multiprocessor (SM).
Each task can include specialized logic, such as warp specialization or tensor core calls.


\textbf{Event Tensor.}
An Event Tensor is a multi-dimensional structure whose elements represent events---the completion of task sets at the SM level---following established practices in parallel programming systems~\cite{blumofe1995cilk,treichler2014realm,bauer2012legion,dagum1998openmp}.
Each element has an initial wait count recording the number of tasks it depends on and supports several operations:
\texttt{E[i].notify()} signals task completion, \texttt{E[i].wait()} blocks until the event is triggered, and in dynamic scheduling, events can also trigger dependent tasks.
Compared with approaches that manage standalone events individually, Event Tensors greatly reduce task-graph management overhead when scaling to millions of fine-grained events in real LLM inference workloads.

\textbf{Graph function.} A graph function represents a computational graph consisting of \calldevice{} calls that explicitly launch device functions with specified task shapes.
Unlike traditional computational graphs, it includes both data tensors and Event Tensors.
Each device function launch can annotate explicit input/output dependencies and coordinate mappings that track fine-grained task relationships through Event Tensors.


Figure~\ref{fig:example_code} shows an example Event Tensor--based program.
The general program can be viewed as a compact representation of task graphs in the form of ``producer task $\rightarrow$ event $\rightarrow$ consumer task''. 
We use generic lambda functions to represent event task relations.\footnote{For simplicity, we use Numpy einsum-like notations~\cite{Harris_2020_numpy} in figures and example codes.}
These dependency annotation implicitly maps to event notifications at the end of each producer task and waiting at the beginning of each consumer task.
We find this notation to be sufficient for most use cases. 
Importantly, we also allow device function to explicitly take in Event Tensors as arguments and call the event notify and wait inside each task. The first-class support of the Event Tensor in device function enables us to represent advanced use cases. It also enables explicit fusion optimizations as transformations within the representation that we will discuss in detail in the next section.



\hiddenNote{How dependency is represented}

\hiddenNote{Row-sum: a concrete example in practice}

\begin{figure}[!t]
    \centering
    \includegraphics[width=0.45\textwidth]{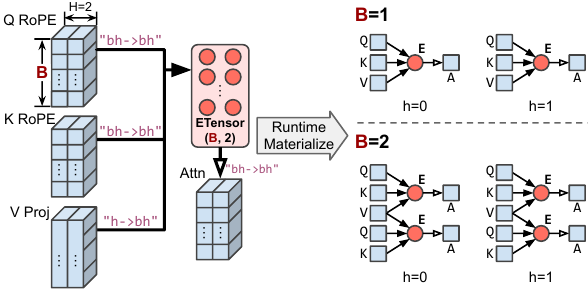} 
    \vspacebeforecap
    \caption{\textbf{Event Tensor handles shape dynamism} with symbolic-shape tensors that define a template for dependency graphs.
    At runtime, the template is instantiated with concrete shape values (e.g., producing a $1 \times 2$ graph for batch size 1 or a $2 \times 2$ graph for batch size 2) without recompilation or repeated graph capture.
    }
    \label{fig:dynamic-materialize} 
\end{figure}

\subsection{Representing Fine-Grained Dependencies}

To illustrate how Event Tensors are used in practice, we walk through the example shown in Figure~\ref{fig:example_code}.
It shows a task graph that performs a summation over the inner axis of the input tensor $A$, which has symbolic shape $(n \times 32, 128)$: $C[i] = \sum_{k \in [0, 128)} A[i, k]$.
The example adopts a split-K algorithm that divides the summation into two stages $B[i, j] = \sum_{k \in [j*32, j*32+32)} A[i, k]$ and $C[i] = \sum_{k \in [0, 4)} B[i, k]$,
where the first stage computes partial sums of each row into $B$, and the second stage aggregates them to produce $C$.
In a traditional kernel-by-kernel approach, tasks in the second stage
are launched only after all tasks in the first stage have completed.
However, each output row $C[i]$ depends only on the corresponding row $B[i, :]$,
meaning that its computation can proceed concurrently with partial sums of other rows.
To capture this fine-grained dependency, we partition the computations of $B$ and $C$ into finer tasks and introduce an Event Tensor $E$:
\begin{align*}
    \text{Task $\hat{B}_{i,j}$:} && B[i*32: i*32+32, j], \\
    \text{Event $E_{i}$:} && E[i],\\
    \text{Task $\hat{C}_{i}$:} && C[i*32:i*32+32].
\end{align*}
With task partitioning, the dependency relations becomes
$\hat{B}_{i, j} \rightarrow E_{i}$ ($\hat{B}_{i, j}$  produces $E_i$) and
$E_{i} \rightarrow \hat{C}_{i}$ ($\hat{C}_i$ consumes $E_i$),
where each $E_{i}$ corresponds to the completion of 32 consecutive rows of $B$ starting from row $32 * i$.

\hiddenNote{Code explanation}

The \texttt{main\_graph} function provides the description of the overall computation where the primitive \calldevice{} first launches  \texttt{partial\_sum} and then \texttt{final\_sum} function. The dependencies between \texttt{partial\_sum} and  \texttt{final\_sum}  tasks are specified through \texttt{out\_edges} and \texttt{in\_edges} arguments.



\hiddenNote{Benefits of the Event Tensor abstraction}

\subsection{First-Class Dynamic Shape Support}

The Event Tensor-based graph can be viewed as a more compact representation of the task graph representations~\cite{treichler2014realm} in parallel systems, where each event element and tasks are explicitly represented as individual nodes and edges materialized as task graphs in the runtime.
In our case, we can use a single tensor to represent thousands of events. The Event Tensor representation also allows us to bring in first-class support for symbolic dimensions. For example, we can have an Event Tensor shape to contain symbolic variables such as batch size of sequence length. The symbolic-shape Event Tensor graph serves as a generic template that corresponds to different task graphs~(Figure~\ref{fig:dynamic-materialize}) at runtime. This powerful representation gives us the ability to overcome limitations in static task-graph systems such as CUDA Graph, and~\emph{ahead of time} optimize dynamic shape Event Tensor graph for multiple shapes without recompilation or repeated graph capture,
providing greater flexibility in handling dynamic shape workloads.

\subsection{Supporting Data-Dependent Dynamism}

\begin{figure}[t]
\centering
\includegraphics[width=0.43\textwidth]{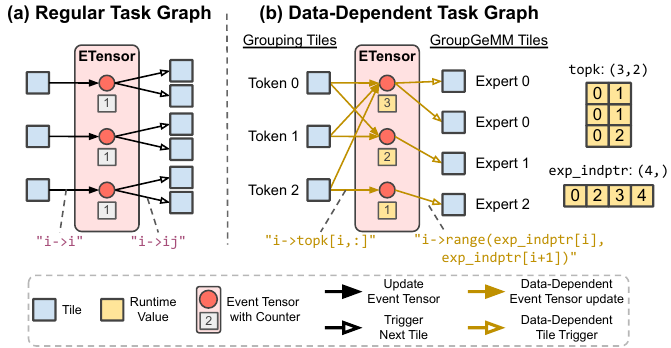}
\caption{\textbf{Event Tensor handles data-dependent dynamism} by
(a) A regular workload with static dependencies. (b) A data-dependent MoE workload where runtime tensors \texttt{topk} and \texttt{exp\_indptr} define an irregular task graph, enabling data-dependent event updates and task triggering.}
\label{fig:data_dependent_event}
\vspaceaftercap
\vspace{-0.5em}
\end{figure}


A critical challenge in modern workloads is handling irregular, data-dependent task graphs. A representative example is the Mixture-of-Experts (MoE) layer~\cite{shazeer2017outrageouslylargeneuralnetworks}, where input tokens are dynamically routed to different expert sub-networks based on routing decisions computed at runtime.
An efficient MoE implementation typically first groups tokens according to their assigned experts and then uses GroupGEMM operators to compute the results for all tokens within each group.

Dynamic routing introduces fine-grained, data-dependent task dependencies that are unknown at compile time (specifically, which expert’s GroupGEMM tile processes which tokens). This dynamism poses a significant challenge for traditional compilers and schedulers, which assume a static task graph with fixed dependencies.
To efficiently represent such dynamic workloads, we need an abstraction that can
(1) determine at runtime which tasks each consumer task depends on, and
(2) trigger a variable number of consumer tasks based on runtime data.
The Event Tensor abstraction is designed precisely for this purpose. It manages data-dependent task graphs through two core mechanisms:

\MyPara{Data-Dependent Event Update.}
Unlike conventional task graphs that support only static dependencies~(Figure~\ref{fig:data_dependent_event}a), the Event Tensor abstraction allows dynamic event dependencies~(Figure~\ref{fig:data_dependent_event}b).
In the MoE example, runtime routing decisions stored in the \texttt{topk} tensor determine which grouping tiles (one per token) update which events (one per expert). Each expert's event counter is initialized to the number of tokens routed to it, and this initialization occurs dynamically at runtime, together with the computation of \texttt{topk}.


\MyPara{Data-Dependent Task Triggering.}
Similarly, an event can trigger a runtime-dependent number of tasks.
Based on routing decisions in \texttt{topk}, we can compute how many tokens each expert must process and, therefore, how many GroupGEMM tiles each expert requires.
As illustrated in Figure~\ref{fig:data_dependent_event}b, this information is encoded in the tensor \texttt{exp\_indptr}, which stores the prefix sum of GroupGEMM tiles to be triggered per expert.\footnote{\texttt{indptr} is a term commonly used in sparse matrix representations such as the compressed sparse row (CSR) format.}
Leveraging \texttt{exp\_indptr}, we enable data-dependent triggering, where expert~\texttt{i} activates tiles in the range \texttt{(exp\_indptr[i], exp\_indptr[i+1])}.


Together, these two mechanisms allow the compiler to generate megakernels that efficiently adapt to highly dynamic workloads---cases that static task graphs handle poorly.
\revised{Combined with the symbolic-shape support described above, all compilation in \sys{} occurs offline; at inference time, the compiled binary handles both shape and data-dependent dynamism with zero compilation overhead.}

\revised{Note that the overall dependency chain remains strictly feed-forward (Figure~\ref{fig:overview}, right): Attention Output $\rightarrow$ Token Routing (TopK) $\rightarrow$ Token Grouping $\rightarrow$ Token Computation (GroupGEMM). Computing TopK depends only on the preceding Attention output---a standard static dependency. The data-dependent Event Tensor mechanisms described above govern only the later stages, where routing results dynamically determine which grouping tasks notify which expert events and how many GroupGEMM tiles each expert triggers.}

\section{Event Tensor Compilation}

This section describes optimizations in \sys{} that make use of the proposed Event Tensor abstractions.

\subsection{Static Scheduling and Transformation}
\label{sec:static-scheduling}


\hiddenNote{What static scheduling is.}

Static scheduling fuses multiple device functions together by explicitly distributing tasks across
streaming multiprocessors (SMs) ahead of time. As a result, each task is pre-assigned to the task queue of a specific SM.
Task dependencies are managed through low-level synchronization primitives, such as counter-based semaphores and event-triggered waits.
This approach achieves minimal synchronization overhead and is particularly effective for predictable workloads where SM partitions can be optimized ahead of time.

\begin{figure*}[!t]
    \centering
    \includegraphics[width=0.90\textwidth]{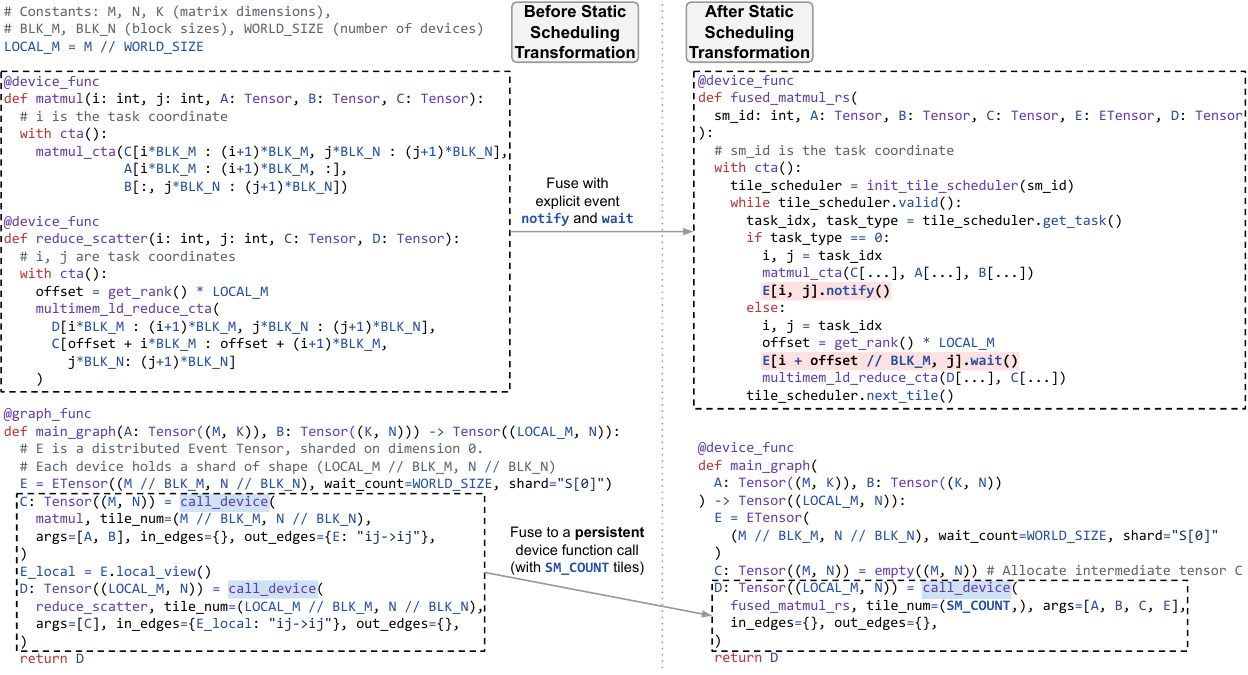} 
    \vspacebeforecap
    \caption{GEMM + Reduce-Scatter before and after static scheduling transformation.
    Two separate device functions are fused into a single persistent function, with explicit \texttt{notify} and \texttt{wait} calls on the Event Tensor to coordinate dependencies.}
    \vspaceaftercap
    \label{fig:static-schedule-code} 
\end{figure*}

\hiddenNote{Static scheduling with Event Tensor. Notify-and-wait}

\hiddenNote{with GeMM+RS example}

\begin{figure}[!t]
    \centering
    \includegraphics[width=0.43\textwidth]{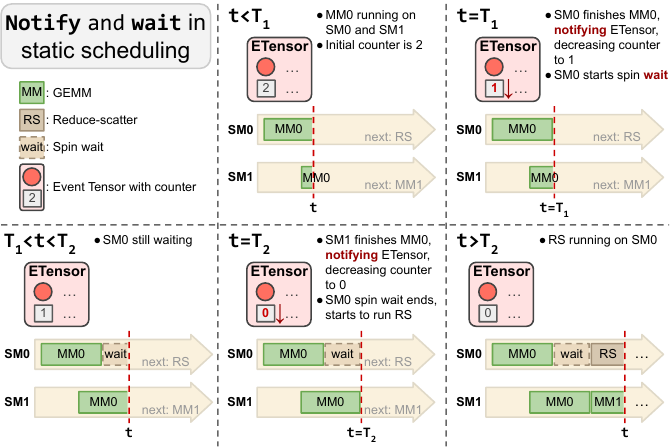} 
    \vspacebeforecap
    \caption{Notify-and-wait mechanism for static scheduling.
 }
    \vspaceaftercap
    \label{fig:static-schedule-runtime-impl} 
\end{figure}

\hiddenNote{We have a static schedule transformation. Before/After}

We implement a static scheduling transformation in \sys{} with three main steps~(Algorithm~\ref{alg:static-schedule}):
(1) construct per-SM execution queues on the host;
(2) generate a persistent main loop that lets each SM execute tasks continuously without relaunching; and
(3) lower Event Tensor dependencies into explicit \texttt{notify()} and \texttt{wait()} calls to enforce fine-grained execution order.
Figure~\ref{fig:static-schedule-code} illustrates this process for GEMM + Reduce-Scatter, fundamental to tensor-parallel execution.
It fuses the device functions into a single persistent kernel, whose main loop continuously fetches tasks from a precomputed queue in \texttt{tile\_scheduler} and issues \texttt{notify()} at the end of GEMM tasks and \texttt{wait()} at the beginning of Reduce-Scatter tasks.

Figure~\ref{fig:static-schedule-runtime-impl} illustrates how the statically fused GEMM (MM) + Reduce-Scatter (RS) kernel operates in practice.
Each RS task depends on two MM tasks (i.e., an RS tile spans twice the size of an MM tile), so the initial counter for each event is two.
At $T_1$, MM0 on SM0 finishes and notifies the Event Tensor, reducing the counter to one.
The RS task, statically scheduled next on SM0, cannot yet proceed and enters a spin-wait state.
Between $T_1$ and $T_2$, SM1 continues executing MM0, keeping the GPU busy.
At $T_2$, MM0 on SM1 completes, decrementing the counter to zero and satisfying the dependency, which releases the RS task from its wait loop and allows execution to begin on SM0.


\begin{algorithm}[tb]
   \scriptsize
   \caption{Static Scheduling Transformation in \sys{}}
   \label{alg:static-schedule}

\begin{algorithmic}[1]
   \STATE {\bfseries Input:} A module \texttt{mod} containing a tile-level dataflow graph \texttt{G} with Event Tensor dependencies.
   \STATE {\bfseries Output:} An updated module with a fused, statically scheduled megakernel.
   \STATE \texttt{mod\_updated} $\leftarrow$ \texttt{mod.Copy()}
   \STATE \texttt{static\_schedule} $\leftarrow$ \texttt{GenerateStaticSchedule(G)}
   \STATE \texttt{fused\_kernel} $\leftarrow$ \texttt{NewPersistentKernel()}
   \STATE // Embed the pre-computed schedule into global memory
   \STATE \texttt{fused\_kernel.AddBuffer(static\_schedule)}
   \FORALL{\texttt{task\_grid} in \texttt{G}}
        \STATE \texttt{fused\_kernel.AddDispatchLogic(task\_grid)}
        \FORALL{\texttt{event} in \texttt{task\_grid.in\_edges}}
            \STATE \texttt{fused\_kernel.AddWaitLogic(event)}
        \ENDFOR
        \STATE \texttt{fused\_kernel.AddTileLogic(task\_grid)}
        \FORALL{\texttt{event} in \texttt{task\_grid.out\_edges}}
            \STATE \texttt{fused\_kernel.AddNotifyLogic(event)}
        \ENDFOR
   \ENDFOR
   \STATE \texttt{mod\_updated.Replace(G, fused\_kernel)}
   \STATE \textbf{return} \texttt{mod\_updated}
\end{algorithmic}
   \normalsize
\end{algorithm}

Plain static scheduling does not naturally support dynamic workloads.
To handle shape dynamism, we sample a set of representative shapes; unseen shapes reuse the execution queue of the next larger sampled value.
To handle data-dependent dynamism, we conservatively assume the worst case for Event Tensor updates and triggers by rewriting related \texttt{notify()} and \texttt{wait()} operations to \texttt{E[0].notify()} and \texttt{E[0].wait()}. For simplicity, we use a round-robin policy to construct execution queues.

\hiddenNote{Benefits and differences to existing work}
\hiddenNote{consider move to related work}

\subsection{Dynamic Scheduling and Transformation}
\label{sec:dynamic-scheduling}

\begin{figure}[!t]
    \centering
    \includegraphics[width=0.42\textwidth]{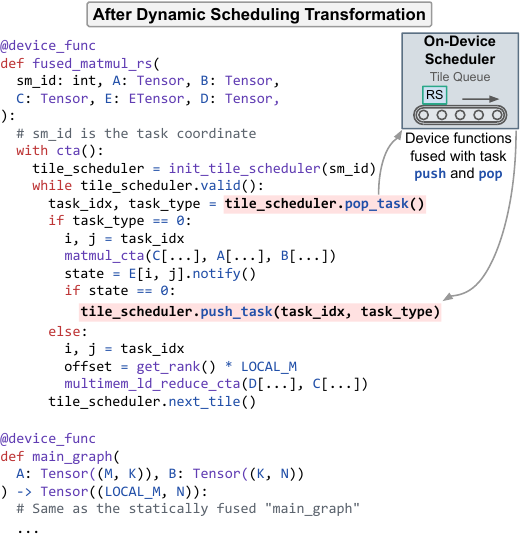} 
    \vspacebeforecap
    \caption{GEMM + Reduce-Scatter after dynamic scheduling transformation.
    Task \texttt{push} and \texttt{pop} are inserted, and task execution is dynamically coordinated by the scheduler.}
    \label{fig:dynamic-schedule-code} 
\end{figure}

\hiddenNote{What dynamic scheduling is.}

\hiddenNote{Dynamic scheduling with Event Tensor. Push-and-pop}

When task execution time is unpredictable, dynamic scheduling improves load balance across SMs.
\sys{} implements Event Tensor--based dynamic scheduling using a lightweight on-GPU task scheduler.
When an event is triggered---after all its dependent tasks complete---it atomically \textbf{pushes} all associated consumer tasks into the scheduler, marking them ready for execution.
Any available SM can then atomically \textbf{pop} a ready task and execute it.

\begin{figure}[t]
    \centering
    \includegraphics[width=0.43\textwidth]{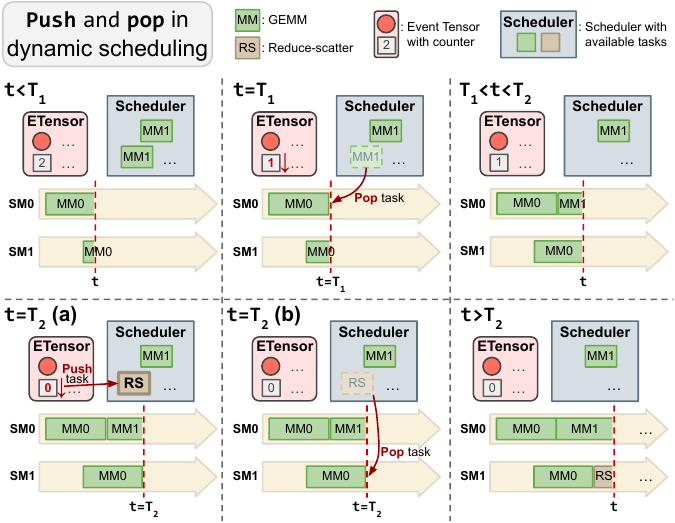} 
    \vspacebeforecap
    \caption{Push-and-pop mechanism for dynamic scheduling.}
    \vspaceaftercap
    \label{fig:dynamic-schedule-runtime-impl} 
\end{figure}

 
\hiddenNote{with GeMM+RS example}

We implement a dynamic scheduling transformation in our compiler\footnote{The pseudocode of the dynamic scheduling transformation is provided in Appendix A.}.
Figure~\ref{fig:dynamic-schedule-code} shows the transformed code for the same GEMM (MM) + Reduce-Scatter (RS) example discussed in \S\ref{sec:static-scheduling}.
Figure~\ref{fig:dynamic-schedule-runtime-impl} illustrates the push–pop mechanism.
At $T_1$, MM0 on SM0 finishes and decrements the event counter to one; SM0, now idle, immediately pops a ready task (MM1) from the scheduler.
At $T_2$, MM0 on SM1 completes, reducing the counter to zero and triggering the RS task to be pushed into the scheduler.
SM1 then pops the RS task and begins execution.
This entire process of dependency tracking and task dispatching occurs efficiently on the GPU, without requiring any host-precomputed task queue.

\hiddenNote{We have a dynamic schedule transformation. Before/After}

Dynamic scheduling inherently supports both kind of dynamism, because tile execution order is decided on the fly at runtime, when the symbolic shape value and runtime value are already resolved by scheduler. Our implementation of push-pop interface uses a centralized queue in global memory shared across all SMs. We choose this design for its implementation simplicity, though we acknowledge potential contention at scale. We also discuss the runtime optimization for dynamic scheduler in Appendix E.

\MyPara{Trade-off between static and dynamic scheduling.}
The choice between static and dynamic scheduling reflects a classic trade-off. Static scheduling leads to minimal scheduling overhead, making it well-suited for predictable workloads. Dynamic scheduling, by contrast, offers flexibility for data-dependent dynamic workload or unpredictable task completion times, naturally achieving load balance at the cost of a small runtime overhead of task queue pushes and pops. 

\subsection{Lowering to Minimal Runtime}

The static and dynamic scheduling in our compiler allows us to encapsulate low-level task dependencies and their handling directly in the transformed program, resulting in minimal needs for corresponding supporting runtime~(Figure~\ref{fig:runtime_arch}). Specifically, each Event Tensor is lowered to an integer tensor, reusing the existing tensor data structure and avoiding any dedicated runtime data structures for events. 
The \texttt{notify()} and \texttt{wait()} operations on this integer tensor are implemented with efficient hardware atomics: \texttt{notify()} performs an atomic decrement, while \texttt{wait()} spin-waits for the counter to reach zero.
Our runtime data state consists solely of these integer tensors and scheduler's task queue. This compiled-in approach brings a smaller runtime requirement compared to a typical task-graph approach, where the entire task graph needs to be materialized in memory and relies on a generic task executor that traverses this graph to launch device functions.




\begin{figure}[!t]
\centering
\includegraphics[width=0.40\textwidth]{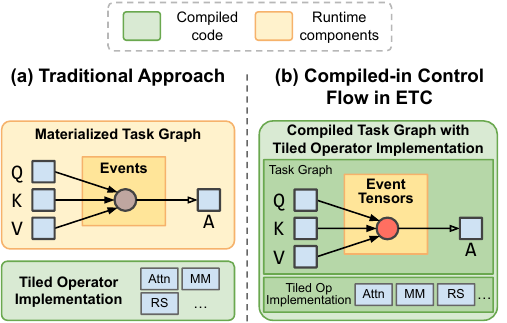} 
\vspacebeforecap
\caption{Comparison of runtime architectures.
(a) The task graph in traditional runtime executor is materialized in memory,
and only tiled operators are compiled.
(b) \sys{} compiles scheduling logic into megakernels
without runtime task graph materialization. 
}
\vspaceaftercap
\label{fig:runtime_arch} 
\end{figure}

\subsection{End-to-End Compilation Flow}

The end-to-end compilation flow of \sys{}\footnote{A compilation pipeline figure can be found in Appendix B.}
starts from an unoptimized computational graph where Event Tensors are defined and operators are already partitioned into CTA-level tiles---either user-specified through a kernel DSL such as Triton~\cite{tillet2019triton} or provided as compiler builtins.
\revised{In our implementation, device functions are written in a TVM-based DSL~\cite{hou2026axesimpleunifiedlayout} that supports standard tile-based programming. Notably, the Event Tensor abstraction is DSL-agnostic: its dependency graph and scheduling logic can be integrated into other compiler stacks (e.g., Triton, CuteDSL) without fundamental design conflicts.}op
The graph first undergoes standard graph-level optimizations, including memory planning, similar to existing deep learning compilers~\cite{xla,lattner2021mlir,lai2025relax,PyTorch2}.
Next, tile-level optimizations refine each operator by determining low-level details such as hardware instruction mapping and pipelining strategies.
The graph is then transformed via either the static or dynamic scheduling pass described in \S\ref{sec:static-scheduling} and \S\ref{sec:dynamic-scheduling}. 
The resulting fused device function is emitted as GPU code in the persistent-kernel style. A subsequent prefetching pass generates weight-prefetching functions for tiled operators based on user annotations, enabling each tile to prefetch weights before input activations arrive. Finally, if static scheduling is chosen, the compiler computes the per-SM task order and materializes it as the megakernel’s static execution queue.

\section{Evaluation}

We implement \sys{} as a series of compiler passes building upon Apache TVM. Notably, our proposed abstraction can be applied to other compilers as well. This section provides evaluations to answer the following key questions:

\squishlist
    \vspace{-0.2em}
    \item How effectively does the Event Tensor abstraction manage fine-grained dependencies for workloads with both static task graphs  (\S \ref{sec:fused_comm}) and task graphs with shape dynamism and data-dependent dynamism (\S \ref{sec:moe_eval})? 
    \vspace{-0.2em}
    \item Do \sys{}-compiled megakernels achieve lower end-to-end latency in dynamic low-batch serving scenarios? (\S\ref{sec:e2e_serving_eval})
    \vspace{-0.2em}
    \item Does the Event Tensor's support for shape dynamism eliminate the significant engine warmup overhead required by static-shape runtime just-in-time (JIT) compilation and graph capture systems? (\S\ref{sec:warmup_overhead})
    \item Do static and dynamic scheduling exhibit distinct performance trade-offs on different workloads? (\S\ref{sec:scheduling_tradeoff})
\squishend

    

\looseness=-1
All experiments are conducted on a server equipped with 8 NVIDIA B200 GPUs connected via NVLink, running Ubuntu 24.04 with \revised{PyTorch 2.8.0,} CUDA 13.0 and driver version 580.82.07. \revised{We choose B200 as it represents state-of-the-art hardware; the Event Tensor abstraction operates at the compiler IR level and is not specific to any particular GPU generation.}
We evaluate \sys{} against state-of-the-art deep learning compilers, specialized libraries, and high-performance LLM serving systems.
Existing megakernel frameworks are tailored to single-batch inference and thus cannot be fairly compared under dynamic-shape or data-dependent workloads.

\subsection{Fused Communication and Computation Performance}
\label{sec:fused_comm}

To evaluate how effectively the Event Tensor abstraction optimizes static compute-communication patterns, we benchmark two fundamental fused kernels for tensor-parallel LLMs: GEMM + Reduce-Scatter and All-Gather + GEMM. These kernels are critical for minimizing latency and maximizing hardware utilization in distributed inference.
We use MLP configurations derived from a range of modern LLMs, fixing the tensor-parallel size to 8 \revised{and the number of tokens to 8192} in all experiments. The configuration details are provided in Appendix C. We compare \sys{}’s generated kernels against several baselines, with implementation choices tailored to each workload’s characteristics:



\squishlist
    \vspace{-0.2em}
    \item GEMM + Reduce-Scatter:
    The Reduce-Scatter collective is implemented using CUDA multimem PTX instructions. We employ \sys{}’s \textbf{dynamic scheduler} to handle unpredictable workloads arising from network contention and fluctuation, where its ability to adapt and balance tasks on the fly is most effective.
    \vspace{-0.2em}
    \item All-Gather + GEMM:
    The All-Gather operation uses a ring algorithm implemented via the copy engine (DMA). We use the \textbf{static scheduler}, as only GEMM tiles execute on SMs following the data arrival order dictated by the ring algorithm. A precomputed static schedule effectively overlaps communication and computation with minimal runtime overhead.
\squishend

The baselines for comparison include:

\squishlist
    \vspace{-0.2em}
    \item cuBLAS+NCCL:  A non-overlapped baseline executing cuBLAS and NCCL kernels sequentially, representing performance without fusion.
    \vspace{-0.2em}
    \item TP-Async~\cite{liang2024torchtitan}: A PyTorch-based approach that manually orchestrates asynchronous operations for overlap.
    \vspace{-0.2em}
    \item Triton Distributed \revised{v0.0.2-rc}~\cite{zheng2025tritondistributedprogrammingoverlappingkernels}: A compiler-based system that generates overlapping kernels, serving as a state-of-the-art open-source baseline.
    \vspace{-0.2em}
    \item cuBLASMp~\cite{nvidia_cublasmp_docs}: A high-performance, multi-process fused-kernel library that overlaps distributed computation and communication.
\squishend


\begin{figure}[!t]
\centering
\includegraphics[width=0.44\textwidth]{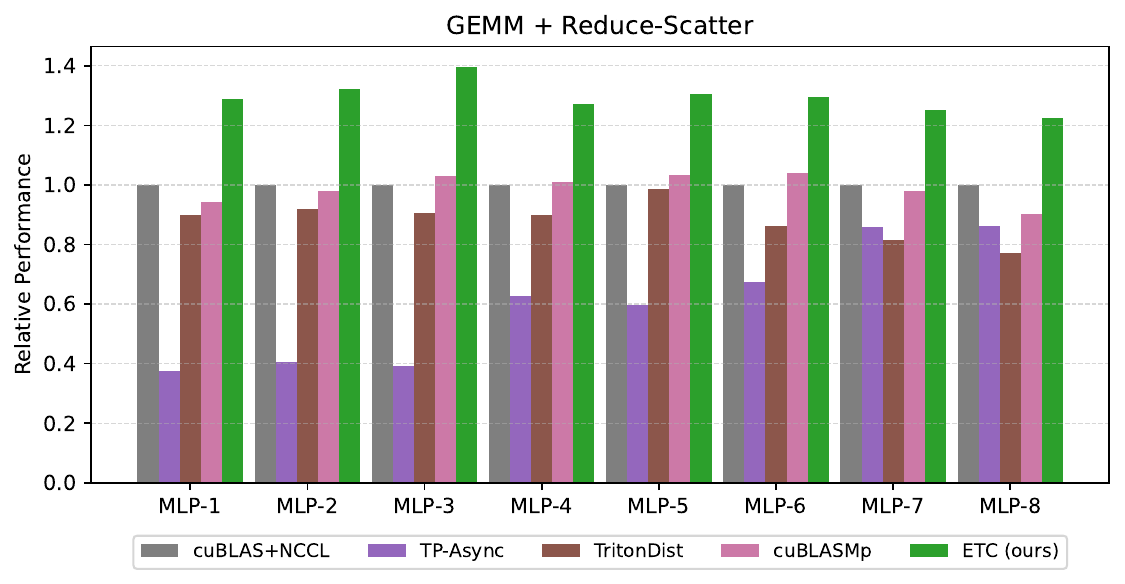} 
\vspacebeforecap
\caption{Performance results of GEMM + Reduce-Scatter on 8 B200s with dynamic scheduler.}
\vspaceaftercap
\label{fig:gemm_rs_result} 
\end{figure}

\begin{figure}[!t]
\centering
\includegraphics[width=0.44\textwidth]{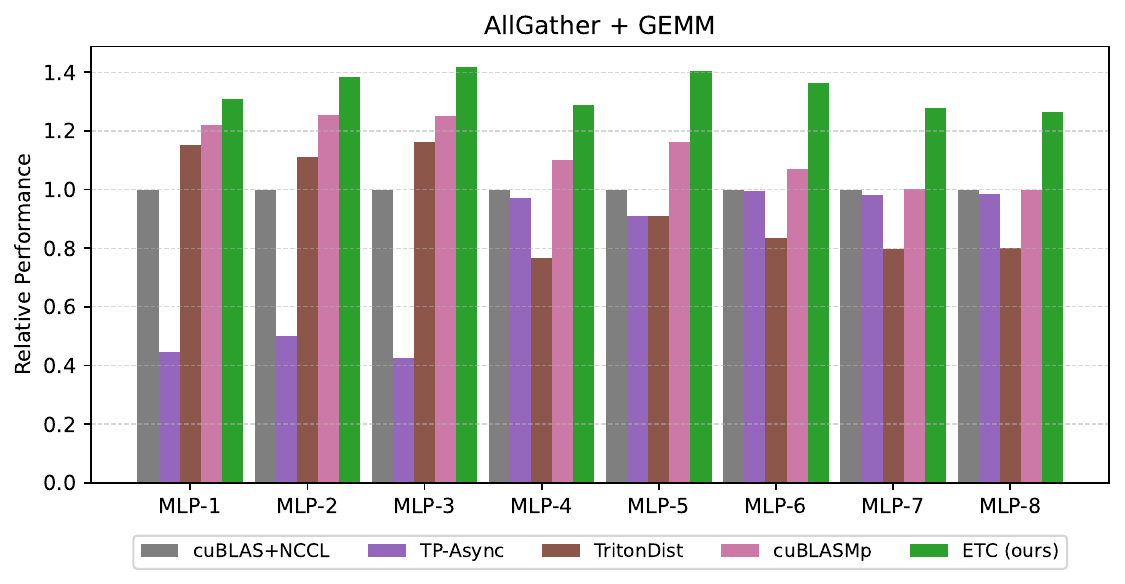} 
\vspacebeforecap
\caption{Performance results of All-Gather + GEMM on 8 B200s with static scheduler.}
\vspaceaftercap
\label{fig:ag_gemm_result} 
\end{figure}

Figure~\ref{fig:gemm_rs_result} and Figure~\ref{fig:ag_gemm_result}
show clear improvements over the baselines,
particularly on larger model configurations, achieving up to a 1.40x execution time speedup over the cuBLAS+NCCL baseline for both workloads.
\revised{Among the fused baselines, TP-Async's coarse-grained splitting can lead to chunks that are either too small to saturate SMs or too large to effectively hide communication latency, and Triton-Dist's experimental B200 support means its Triton-based GEMM is not yet fully optimized for the Blackwell architecture. As a result, the unfused cuBLAS+NCCL baseline is sometimes competitive with these fused approaches, underscoring the difficulty of achieving efficient fusion.}
The consistent performance advantage of \sys{} stems from the Event Tensor abstraction. By representing fine-grained dependencies as a first-class Event Tensor, our compiler can transform monolithic operations into a deeply pipelined task graph. This abstraction also allows our unified scheduling transformations to be applied effectively: we use the dynamic scheduler for GEMM + Reduce-Scatter to handle any potential unpredictability of communication latency, and the static scheduler for All-Gather + GEMM to orchestrate overlap with the predictable ring algorithm with minimal overhead. This fine-grained, compiler-driven approach keeps both compute (SMs) and network resources continuously busy, achieving a degree of overlap that matches or exceeds existing systems.

\subsection{Mixture-of-Experts (MoE) Layer Performance}
\label{sec:moe_eval}
To evaluate our Event Tensor abstraction's ability to manage task graphs with shape dynamism and data-dependent dynamism, this subsection benchmarks a complete MoE layer with variable number of tokens. While existing systems use efficient persistent GroupGEMM kernels, they still require a sequence of separate launches for a complete MoE layer.
\sys{} allows us to fuse the entire data-dependent MoE dataflow into a single megakernel,
and we evaluate its performance against existing optimized multi-kernel baselines.
We benchmark a complete MoE layer in Qwen3-30B-A3B, which has 128 experts with a top-k of 8. The workload consists of processing a variable number of input tokens. For this workload, we use \sys{}'s dynamic scheduler, as its adaptive load balancing is ideal for the irregular, data-dependent task graph. We compare against several baselines:

\squishlist
    \vspace{-0.2em}
    \item Triton \revised{3.4.0}~\cite{tillet2019triton}: A highly optimized MoE implementation widely used in state-of-the-art serving systems, including SGLang~\cite{zheng2024sglang} and vLLM~\cite{10.1145/3600006.3613165}.
    \vspace{-0.2em}
    \item FlashInfer \revised{0.2.14.post1}~\cite{ye2025flashinfer}: A high-performance library providing optimized kernels for LLM inference\revised{, including fused MoE kernels}.
\squishend



\begin{figure}[!t]
\centering
\includegraphics[width=0.44\textwidth]{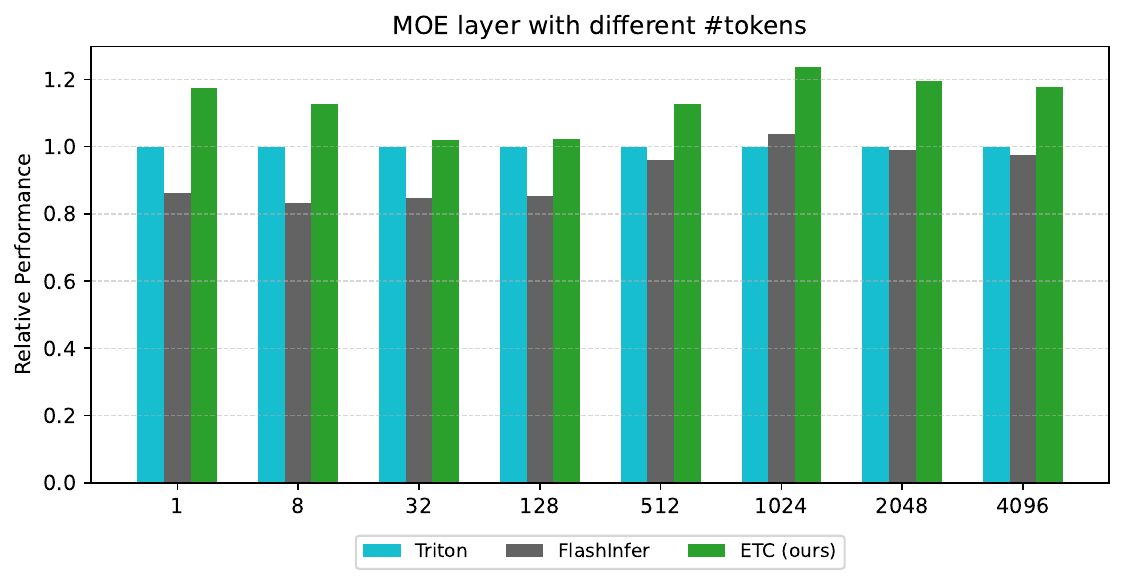} 
\vspacebeforecap
\caption{Performance results of MoE layer on a single B200.}
\vspaceaftercap
\label{fig:moe_result} 
\end{figure}

Figure~\ref{fig:moe_result} plots the relative end-to-end performance for the MoE layer across different token counts. \sys{}'s megakernel approach significantly outperform the best baseline, achieving up to a 1.23x speedup at 1024 tokens.
\revised{Between the two baselines, FlashInfer's GroupGEMM is more optimized for larger token counts, while Triton benefits from fusing gather/scatter into GroupGEMM; their relative ranking thus varies with token count.}
The performance gains of \sys{} are a direct result of the Event Tensor abstraction and dynamic scheduling transformation. Firstly, data-dependent Event Tensors break the global synchronization barrier in the baselines, creating a fine-grained pipeline between the two GroupGEMM stages in MoE\revised{, which also reduces wave quantization by smoothing out SM allocation across fused operators}. Secondly, and more importantly for MoE, our on-chip dynamic scheduler provides superior load balancing for the irregular token routing, minimizing SM idle time and consistently surpassing all other methods as token counts grow.




\subsection{End-to-End Low-Batch Serving Performance}
\label{sec:e2e_serving_eval}

\begin{figure*}
\centering
\includegraphics[width=0.80\textwidth]{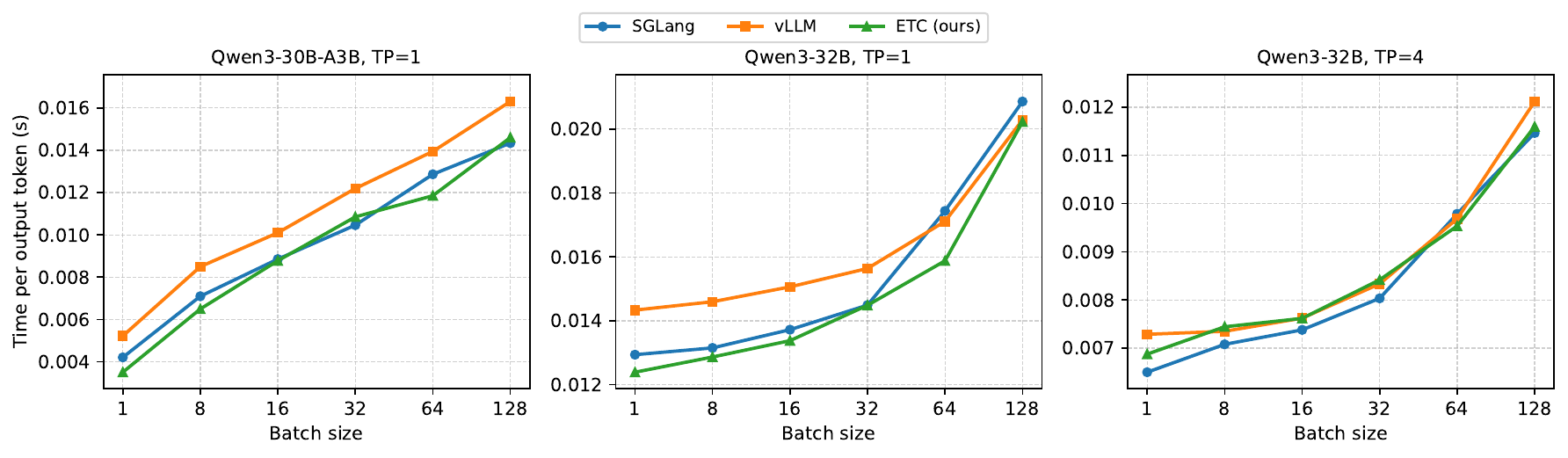} 
\vspacebeforecap
\caption{End-to-end performance of model serving on Qwen3-30B-A3B and Qwen3-32B (Lower is better).}
\vspaceaftercap
\label{fig:eval_e2e_serving} 
\end{figure*}


\looseness=-1
\revised{This subsection tests whether \sys{}-compiled megakernels achieve lower end-to-end latency in dynamic low-batch serving scenarios, which are increasingly dominant for latency-sensitive applications such as real-time agentic workflows and interactive coding assistants.} We focus on the decoding stage, as prefilling and large-batch serving usually have high GPU utilization and benefit marginally from megakernels (except for computation-communication overlap as we showcased in \S\ref{sec:fused_comm}).
\revised{Importantly, \sys{} does not degrade large-batch performance, as hardware utilization of individual operators remains unchanged without extra overhead.}
We integrate \sys{}-compiled megakernels for two representative models: the Qwen3-30B-A3B MoE model and the Qwen3-32B dense model, using the static scheduler for both to avoid runtime overhead.
\revised{The compiled megakernels cover the full decoding pipeline (Attention, RoPE, KV-Cache, Norm, MLP, MoE), not just GEMM. We chose the Qwen family (both dense and MoE variants) as it is architecturally representative of modern LLMs (e.g., LLaMA~3, GPT).}
The benchmark uses a synthetic dataset with a prefill length of 512 and generates 100 output tokens, with batch size varying from 1 to 128. We measure the time-per-output-token (TPOT) metric, which best reflects the LLM engine decoding performance. The raw kernel running time of one decoding iteration can be found in Appendix D, which concentrates only on the kernel performance and excludes all irrelative overhead (e.g., framework-specific scheduling latencies and other CPU overhead),
ensuring a fair comparison of different frameworks.
We compare \sys{} against leading serving systems vLLM \revised{(v0.11.0rc2)} and SGLang \revised{(v0.5.3rc0)} both of which use CUDA Graph and torch.compile~\cite{PyTorch2} for performance optimization.

Figure~\ref{fig:eval_e2e_serving}~(left) shows the end-to-end performance for Qwen3-30B-A3B, where \sys{} achieves a 1.48x speedup over vLLM and 1.20x over SGLang at batch size 1.
In Figure~\ref{fig:eval_e2e_serving}~(mid), for Qwen3-32B, \sys{} consistently delivers the lowest latency, outperforming vLLM by up to 1.15x at batch size 1, and SGLang by up to 1.09x at batch size 64.
In four-way TP tensor-parallel (TP) execution~\cite{shoeybi2020megatronlmtrainingmultibillionparameter} for the Qwen3-32B (Figure~\ref{fig:eval_e2e_serving}, right), \sys{} matches performance of vLLM with a speedup varying between 0.99x and 1.06x. \revised{The latency of both \sys{} and vLLM is higher than SGLang in this setting because SGLang's highly optimized CPU scheduler incurs lower distributed runtime overhead. The occasional small gaps where \sys{} trails the best baseline are attributable to engineering factors---specifically, compiler-generated GEMM tiles that are less tuned than cuBLAS in certain configurations and higher CPU-side overhead in our serving engine---rather than fundamental limitations of the abstraction.}

\looseness=-1
The strong performance of \sys{} stems from its fused megakernel architecture, enabled by the Event Tensor abstraction and static scheduling transformation.
Unlike conventional approaches that launch a sequence of kernels with implicit synchronization at kernel boundaries, \sys{} executes the entire workload within a single persistent kernel. 
This compiler-driven fusion enables fine-grained optimizations that are difficult for CUDA-Graph-based systems: it exposes parallel execution in attention \revised{(e.g., Q's Norm+RoPE running concurrently with K's Norm+RoPE+CacheAppend)}, pipelines GroupGEMMs in MoE and GEMMs in MLP \revised{to reduce wave quantization}, and prefetches model weights before input activations are ready to hide memory latency.
\sys{}'s ability to pipeline and overlap operations across operator boundaries is the key for the latency reduction compared to baselines. Crucially, \sys{} breaks kernel boundaries, successfully applying performance on par with CUDA Graphs to inherently dynamic workloads like MoE.



\subsection{Warmup Overhead}
\label{sec:warmup_overhead}

\begin{table}[!t]
\centering
\small
\vspacebeforecap
\caption{Warmup time of Qwen3-32B model serving using different graph capturing methods.}
\begin{tabular}{ccc}
\toprule
\textbf{Method}       & \multicolumn{1}{l}{\textbf{Warmup Time (s)}} & \multicolumn{1}{l}{\textbf{\# JIT Graph Capture}} \\ \midrule
SGLang (JIT) & 583                                 & 51                                       \\
vLLM (JIT)   & 123                                 & 67                                       \\
\sys{} (AOT)    & 35                                  & 0                                     
\\ \bottomrule

\end{tabular}
\label{tab:warm_up}
\end{table}

This subsection evaluates the deployment impact of \sys{}’s compilation strategy by measuring LLM engine warmup overhead.
\revised{We define warmup time as the total wall-clock time from engine launch to the first request served, including engine initialization, model loading, and all JIT compilation or CUDA Graph capture overheads.}
We test whether \sys{}’s ahead-of-time (AOT) compilation, enabled by shape dynamism support, can eliminate the runtime cost of just-in-time (JIT) compilation and CUDA Graph capture.
Table~\ref{tab:warm_up} shows substantial difference: vLLM requires 123 s and SGLang 583 s to warm up on Qwen3-32B, whereas \sys{} initializes in only 35 s.
This speedup stems from the Event Tensor abstraction, whose first-class shape dynamism support enables AOT compilation.
While baselines must capture many static CUDA Graphs at runtime to cover different shapes (e.g., 67 for vLLM), \sys{} compiles a single persistent, shape-generic megakernel offline (107 s for Qwen3-32B).
At runtime, it simply loads the precompiled graph, avoiding the repetitive warmup penalty inherent to JIT and runtime-capture approaches.



\subsection{Tradeoff Between Different Scheduling Methods}
\label{sec:scheduling_tradeoff}

\begin{table}[!t]
\centering
\small
\vspacebeforecap
\caption{Relative performance of different ETC scheduling methods on MoE
layer against unfused megakernel. \revised{Higher is better.}}
\begin{tabular}{ccccc}
\toprule
\textbf{Num tokens}  & \textbf{1} & \textbf{128} & \textbf{1024} & \textbf{4096} \\ \midrule
Static   & 1.03 & 1.02 &   1.04 &  1.02                                     \\
Dynamic    & 0.95 & 1.06 & 1.08 & 1.03                         
\\ \bottomrule

\end{tabular}
\label{tab:ablation_moe}
\vspaceaftercap
\end{table}

\begin{table}[!t]
\centering
\small
\caption{Relative performance of different ETC scheduling methods on Qwen-3-32B with TP=4 against unfused megakernel. \revised{Higher is better.}}
\begin{tabular}{ccccc}
\toprule
\textbf{Batch Size}  & \textbf{1} & \textbf{16} & \textbf{32} & \textbf{128} \\ \midrule
Static   & 1.09 & 1.06 &   1.07 &  1.06                                     \\
Dynamic    & 0.83 & 0.82 & 0.85 & 0.89                         
\\ \bottomrule

\end{tabular}
\label{tab:ablation_qwen32b}
\vspaceaftercap
\end{table}



This section analyzes the performance characteristics and trade-offs of \sys{}'s static and dynamic scheduling strategies, and quantifies the gains from fusion by comparing them to an unfused megakernel baseline. This baseline uses a single event between different operator stages to enforce a global synchronization barrier, simulating a sequential execution model within a single kernel launch.
\revised{Crucially, the unfused baseline uses identical operator code as \sys{}, so the speedups reported in Tables~\ref{tab:ablation_moe} and~\ref{tab:ablation_qwen32b} are driven purely by the inter-kernel parallelism unlocked by \sys{}'s fine-grained Event Tensor dependencies---the same sources of gain discussed in \S\ref{sec:moe_eval} and \S\ref{sec:e2e_serving_eval} (reduced wave quantization, weight prefetching, parallel execution, and on-chip load balancing)---rather than better operator-level implementations.}

\MyPara{Data-Dependent Workloads.} For workloads with data-dependent control flow, such as the MoE layer in \S5.2, the dynamic scheduler provides load balancing. As shown in Table~\ref{tab:ablation_moe}, dynamic scheduler outperforms static scheduler except on single-batch inference, with the largest speedup being 4.0\% over static scheduler and 8.1\% over unfused baseline when batch size is 1024. The data-dependent routing of tokens creates an inherent workload imbalance. A rigid static scheduler can cause some SMs to accumulate a queue of longer-running tiles, forcing them to become stragglers while other SMs sit idle. 

\MyPara{Regular Workloads.}
Conversely, for the regular, dense transformer layer workload (analyzed with TP=4 in \S5.3), the static scheduler is the clear winner (Table \ref{tab:ablation_qwen32b}). The dynamic scheduler's overhead becomes very large on distributed setting, especially when trying to push tasks to remote task queue. There is also a consistent 6-8\% speedup of \sys{}-static over the \sys{}-unfused version, a gain purely from fine-grained pipelining.

\revised{These results also explain why the multi-GPU evaluation results in Figures~\ref{fig:gemm_rs_result}/\ref{fig:ag_gemm_result} and Figure~\ref{fig:eval_e2e_serving} appear to show different trends. Figures~\ref{fig:gemm_rs_result} and~\ref{fig:ag_gemm_result} represent a bandwidth-bound regime with large batches (8192 tokens), where \sys{} excels by utilizing fine-grained signaling to overlap communication and computation. In contrast, Figure~\ref{fig:eval_e2e_serving} (right) represents a latency-critical regime with low batches, where communication overhead and CPU scheduling overhead are exposed. This necessitates different scheduling strategies: dynamic scheduling handles large-batch jitter effectively, while static scheduling minimizes overhead for latency-sensitive low-batch tasks.}

These findings demonstrate a clear, workload-dependent trade-off, confirming the value of supporting both scheduling transformations within the \sys{} framework.
\section{Related Work}

\hiddenNote{Deep Learning Compilers}
Deep learning compilers such as MLIR~\cite{lattner2021mlir}, XLA~\cite{xla}, TVM~\cite{chen2018tvm,feng2023tensorir,lai2025relax}, and the PyTorch compiler~\cite{PyTorch2} have laid the foundation for optimizing deep learning models.  These systems perform graph-level optimizations and run execution kernel-by-kernel. 
CUDA Graph~\cite{Cudagraph} provides a way to drastically reduce launch overhead by capturing and replaying a sequence of kernels, but relies on static input. Machine learning compilers are also developing 
vertical fusion~\cite{zheng2020ansor,dnnfusion} and horizontal fusion~\cite{jia2019taso, li2022automatic} to optimize kernel launch overhead.
Rammer~\cite{rammer} and Roller~\cite{roller} perform software launch of tile-based tasks. All the previous works do not have explicit abstraction for fine-grained dependencies tracking and optimizations, and can benefit from our proposed Event Tensor to enable megakernel optimizations. \revised{Dynamic tensor compilers such as DynaTune~\cite{zhang2021dynatune}, DietCode~\cite{dietcode}, and SparseTIR~\cite{ye2023sparsetir} handle dynamic shapes or sparsity at the single-kernel level; \sys{} complements them by fusing their operator implementations into megakernels to unlock inter-kernel parallelism.}

\hiddenNote{LLM Inference Frameworks and MegaKernels}
LLM inference systems such as SGLang~\cite{zheng2024sglang}, vLLM~\cite{10.1145/3600006.3613165}, TensorRT-LLM~\cite{tensorrt_llm}, and Orca~\cite{yu2022orca} achieve high performance through system optimizations such as continuous batching and speculative execution. Our event tensor compiler can serve as a backend for these frameworks to enable more efficient GPU execution. Recent works~\cite{mpk, tkmega} start to build megakernels for LLMs. These approaches only supports single-batch dense model inference, and focus on a single scheduling strategy. The event tensor abstraction proposed in this paper complement these approaches by providing systematic compiler abstraction support for shape dynamism and data-dependent dynamism. The event tensor compiler also supports both static scheduling and dynamic scheduling.
\revised{CuSync~\cite{cusync} optimizes co-scheduling of separate kernels on distinct CUDA streams, and FlashMoE~\cite{flashmoe} provides a hand-optimized kernel for distributed MoE. \sys{} differs from both by fusing entire subgraphs into a single persistent megakernel via a systematic compiler pipeline, generalizing beyond any single operator pattern.}

\hiddenNote{Parallel Task Scheduling}

Our approach is closely related to task-based parallel programming models such as Cilk~\cite{blumofe1995cilk}, Legion~\cite{bauer2012legion}, Realm~\cite{treichler2014realm}, and OpenMP Tasks~\cite{dagum1998openmp}. Most of the previous approaches focus on coarse-grained tasks typically orchestrated by the CPU. Our approach is also related to Graphene~\cite{Graphene} and Cypress~\cite{cypress} that optimizes a single kernel. \revised{Graphene models threads as a tensor with synchronization capabilities, which is conceptually related; however, it targets single-kernel optimization rather than multi-operator megakernel fusion with dynamic shape and data-dependent support.} Our approach builds on these previous insights and proposes the event tensor abstraction that compactly represents fine-grained dependencies across operator sub-tasks and runs both static and dynamic scheduling in the GPU streaming multiprocessors.

\section{Conclusion}

This work introduces Event Tensor, a unified abstraction that expresses fine-grained synchronization for compiling dynamic GPU megakernels.
Event Tensor provides first-class support for both shape and data-dependent dynamism.
Built on this abstraction, \sys{} systematically generates high-performance persistent kernels using static and dynamic scheduling.
\sys{} achieves state-of-the-art serving latency while substantially reducing warmup overhead.
\revised{In the future, we envision higher-level passes that automatically generate Event Tensor task graphs from standard computational graphs, further reducing manual synchronization effort.}
We hope this work will encourage additional studies of megakernels and highlight new possibilities for ML compilers.

\section*{Acknowledgements}
We thank all anonymous MLSys reviewers and our shepherd for their constructive feedback and comments. This work is supported in part by gifts from NVIDIA, Google and Amazon. We also acknowledge support from NVIDIA for the DGX B200.

\bibliography{references}
\bibliographystyle{mlsys2025}

\clearpage
\appendix
\section{Dynamic Scheduling Pseudocode}

\begin{algorithm}[tb]
   \scriptsize
   \caption{Dynamic Scheduling Transformation in \sys{}}
   \label{alg:dynamic-schedule}

\begin{algorithmic}[1]
   \STATE {\bfseries Input:} A module \texttt{mod} containing a tile-level dataflow graph \texttt{G} with Event Tensor dependencies..
   \STATE {\bfseries Output:} An updated module with a fused, dynamic scheduled megakernel.
   \STATE \texttt{mod\_updated} $\leftarrow$ \texttt{mod.Copy()}
   \STATE \texttt{fused\_kernel} $\leftarrow$ \texttt{NewPersistentKernel()}
   \STATE // Not runtime scheduler. Only provides push/pop functions
   \STATE \texttt{scheduler} $\leftarrow$ \texttt{GPUScheduler()} 
   \STATE \texttt{fused\_kernel.AddPopLogic(scheduler.f\_pop\_tasks)}
   \FORALL{\texttt{task\_grid} in \texttt{G}}
        \STATE \texttt{fused\_kernel.AddDispatchLogic(task\_grid)}
        \STATE \texttt{fused\_kernel.AddTileLogic(task\_grid)}
        \FORALL{\texttt{event} in \texttt{task\_grid.out\_edges}}
            \STATE \texttt{fused\_kernel.AddCompleteOnLogic(event, scheduler.f\_push\_tasks)}
        \ENDFOR
   \ENDFOR
   \STATE \texttt{mod\_updated.Replace(G, fused\_kernel)}
   \STATE \textbf{return} \texttt{mod\_updated}
\end{algorithmic}
   \normalsize
\end{algorithm}

\hiddenNote{Pseudo code of dynamic schedule transformation} Algorithm~\ref{alg:dynamic-schedule} describes the compiler pass to transform an event tensor graph to a dynamically scheduled megakernel.
A call to \texttt{scheduler.pop\_tasks} is inserted whenever an SM finishes its current task, while a call to \texttt{scheduler.push\_tasks} is inserted when the completion of a task decrements the associated event counters to zero, thereby unblocking dependent tasks.

\section{\sys{} End-to-End Compilation Flow}

\begin{figure}[t]
\centering
\includegraphics[width=0.40\textwidth]{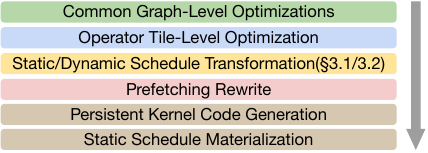}
\caption{End-to-end compilation pipeline in \sys{}.}
\label{fig:e2e_compilation_flow}
\end{figure}

Figure~\ref{fig:e2e_compilation_flow} summarizes the end-to-end compilation flow of \sys{}, as described in \S3.4.

\section{MLP configuration used in \S4.1}
Table~\ref{tab:model-configs} shows MLP configurations used in fused communication and computation evaluation, which are derived from a range of modern LLMs.

\begin{table}[t]
  \centering
  \small
  \setlength{\tabcolsep}{8pt}
  \caption{Model configurations for MLP, where S = sequence length, H = hidden dim, I = intermediate size.}
  \renewcommand{\arraystretch}{1.15}

  \begin{tabular}{lcccl}
    \toprule
    \multicolumn{5}{c}{\textbf{MLP Configurations}}\\
    \midrule
    \textbf{Name} & \textbf{Source Model} & \textbf{S} & \textbf{H} & \textbf{I} \\
    \midrule
    MLP-1 & Qwen3-8B & 8192 & 4096 & 12288  \\
    MLP-2 & LLaMA-3.1-8B & 8192 & 4096 & 14336 \\
    MLP-3 & Gemma-2-9B & 8192 & 3584 & 14336 \\
    MLP-4 & Gemma-2-27B & 8192 & 4608 & 36864 \\
    MLP-5 & Qwen3-32B & 8192 & 5120 & 25600 \\
    MLP-6 & LLaMA-3.1-70B & 8192 & 8192 & 28672 \\
    MLP-7 & GPT-3-175B & 8192 & 12288 & 49152 \\
    MLP-8 & LLaMA-3.1-405B & 8192 & 16384 & 53248 \\
    \bottomrule
  \end{tabular}

  \vspace{1.2ex}

  \label{tab:model-configs}

\end{table}

\section{Raw Kernel Time Evaluation in End-to-End LLM Serving}

\begin{figure}[!t]
\centering
\includegraphics[width=0.49\textwidth]{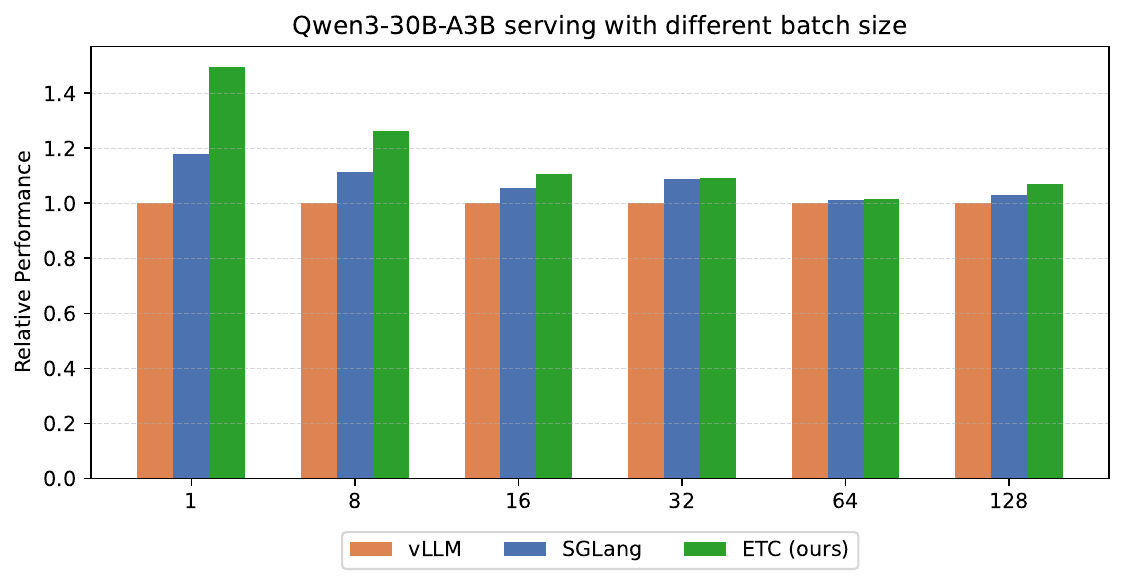} 
\caption{Raw kernel relative performance results of Qwen-30B-A3B on a single B200.}
\label{fig:eval_raw_kernel_qwen30b_a3b_tp1} 
\end{figure}

\begin{figure}[!t]
\centering
\includegraphics[width=0.49\textwidth]{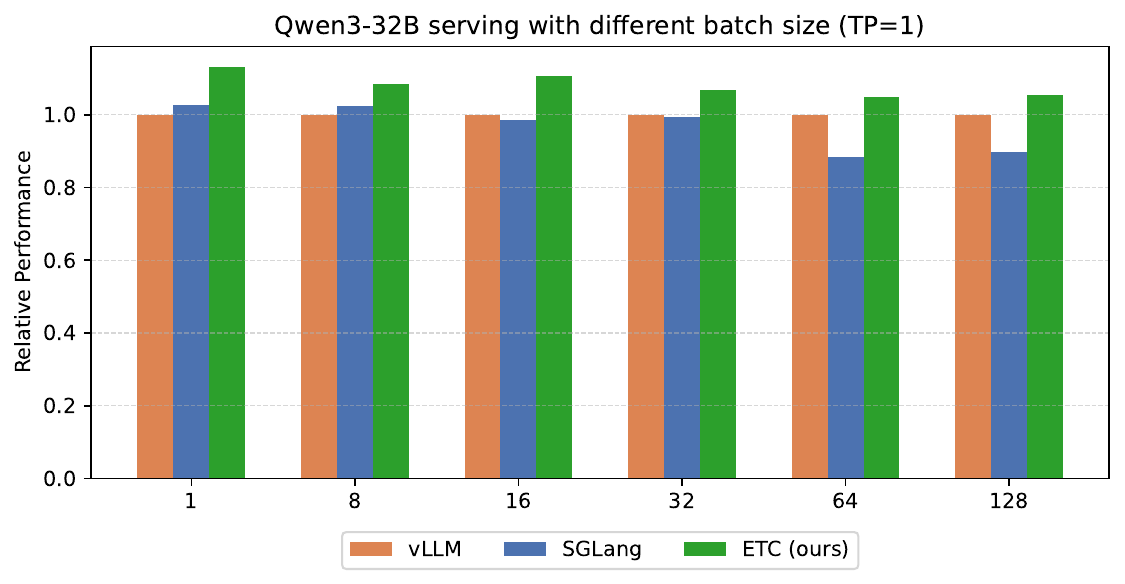} 
\caption{Raw kernel relative performance results of Qwen-32B on a single B200.}
\label{fig:eval_raw_kernel_qwen32b_tp1} 
\end{figure}

\begin{figure}[!t]
\centering
\includegraphics[width=0.49\textwidth]{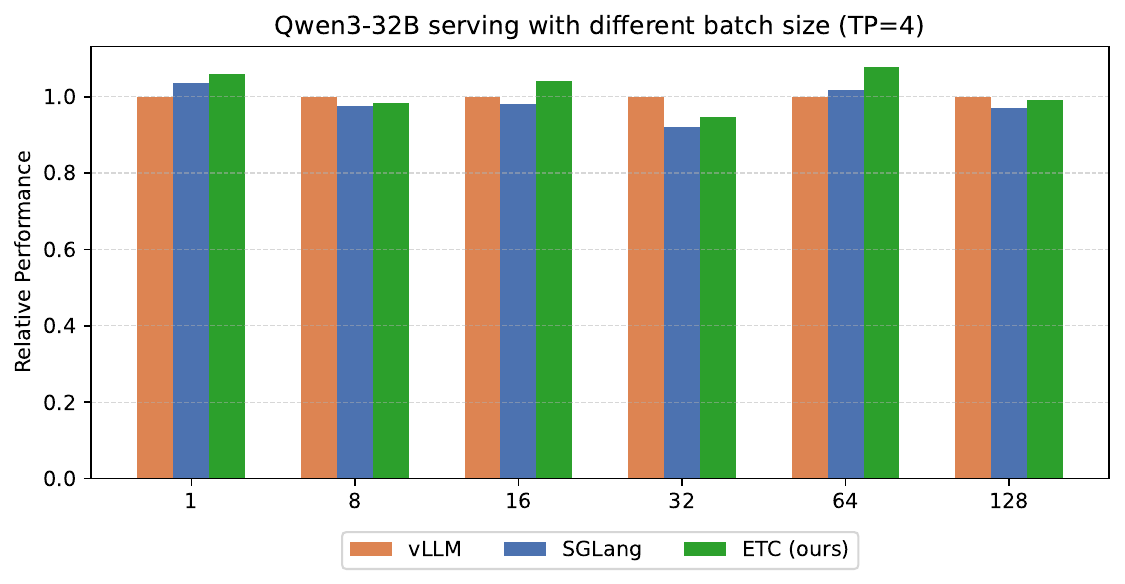} 
\caption{Raw kernel relative performance results of Qwen-32B on four B200s with tensor parallelism.}
\label{fig:eval_raw_kernel_qwen32b_tp4} 
\end{figure}

\S5.3 reports the time-per-output-token (TPOT) metric of \sys{} and baselines in end-to-end LLM serving.
The TPOT number includes not only the total GPU kernel execution time but also CPU-side overheads, such as framework-specific request scheduling latency.
To provide a more direct and fair comparison unaffected by unrelated overheads, this section evaluates the raw GPU kernel execution time in end-to-end LLM serving, using the same baselines and experimental settings as in \S5.3.

Figure~\ref{fig:eval_raw_kernel_qwen30b_a3b_tp1} shows the relative raw kernel end-to-end performance of \sys{} and baselines across multiple batch sizes on Qwen3-30B-A3B, where \sys{} achieves consistent speedups over the baselines, with the most significant improvement being 1.49x over vLLM and 1.27x over SGLang at batch size 1.
With the Event Tensor abstraction supporting data-dependent dependencies in MoE, \sys{} executes the entire MoE model within a single kernel, enabling optimizations such as increased parallelism across attention operators, fine-grained pipelining between GroupGEMMs, and model-weight prefetching.

Figure~\ref{fig:eval_raw_kernel_qwen32b_tp1} and Figure~\ref{fig:eval_raw_kernel_qwen32b_tp4} present the raw kernel performance of Qwen3-32B serving on a single B200 and on four B200s with tensor parallelism, respectively.
Under the single-GPU setting (TP = 1), \sys{} achieves consistent gains over both vLLM and SGLang across all batch sizes, with up to a 1.13x speedup over vLLM at batch size 1 and an average improvement of about 7\%.
In the tensor-parallel case (TP = 4), \sys{} maintains comparable or better performance across all settings, achieving up to a 1.08x speedup over vLLM while sustaining similar scalability as batch size increases.
This on-par performance reflects the effectiveness of \sys{}'s megakernel design and static scheduling based on the Event Tensor abstraction.


\section{Dynamic Scheduler Runtime Optimization}
We adopt an early push strategy to hide scheduling overhead. Rather than waiting for a task's dependencies (its producer tasks) to finish executing, the scheduler proactively pushes a consumer task into the ready queue as soon as all of its producer tasks have been dispatched to SMs (not requiring task completion), and the dependency is guaranteed by the extra wait prior to the consumer's execution. This proactive measure prevents the overhead of the push operation from falling onto the critical path. The push is performed concurrently with the execution of the producer tasks, effectively overlapping the scheduling cost with the preceding computation.


\end{document}